\documentclass[reprint,
aps,pra,    
10pt,
superscriptaddress,
amsmath,amssymb,
floatfix
]{revtex4-1} 

\usepackage{graphicx}
\usepackage{amsmath,amssymb}
\usepackage{mathtools}
\usepackage{dcolumn}
\usepackage{esvect}
\usepackage{booktabs}
\usepackage{titlesec}
\usepackage{siunitx}
\usepackage{verbatim}
\usepackage[svgnames]{xcolor}
\usepackage[colorlinks=true,
linkcolor=MediumBlue,
urlcolor=MediumBlue,
citecolor=MediumBlue]{hyperref}
\usepackage{ulem}
\DeclareMathAlphabet\mathbfcal{OMS}{cmsy}{b}{n}

\titlespacing\section{0pt}{15pt plus 2pt minus 2pt}{6pt plus 2pt minus 2pt}
\titlespacing\subsection{0pt}{15pt plus 2pt minus 2pt}{6pt plus 2pt minus 2pt}

\newcommand{\simsym}{\mathord\sim}

\renewcommand{\vec}[1]{\ensuremath{\mathbf{#1}}}

\begin{document}

\title{Electrically Driven, Optically Levitated Microscopic Rotors}

\author{Alexander D. Rider}
\email{arider@stanford.edu}
\affiliation{Department of Physics, Stanford \; University, Stanford, California \; 94305, USA}

\author{Charles P. Blakemore}
\affiliation{Department of Physics, Stanford \; University, Stanford, California \; 94305, USA}

\author{Akio Kawasaki}
\affiliation{Department of Physics, Stanford \; University, Stanford, California \; 94305, USA}
\affiliation{W. W. Hansen Experimental Physics Laboratory, Stanford \; University, Stanford, California \; 94305, USA}

\author{Nadav Priel}
\affiliation{Department of Physics, Stanford \; University, Stanford, California \; 94305, USA}

\author{Sandip Roy}
\affiliation{Department of Physics, Stanford \; University, Stanford, California \; 94305, USA}

\author{Giorgio Gratta}
\affiliation{Department of Physics, Stanford \; University, Stanford, California \; 94305, USA}
\affiliation{W. W. Hansen Experimental Physics Laboratory, Stanford \; University, Stanford, California \; 94305, USA}

\date{\today}
\begin{abstract}

We report on the electrically driven rotation of 2.4-$\mu$m-radius, optically levitated dielectric microspheres. Electric fields are used to apply torques to a microsphere's permanent electric dipole moment, while angular displacement is measured by detecting the change in polarization state of light transmitted through the microsphere (MS). This technique enables greater control than previously achieved with purely optical means because the direction and magnitude of the electric torque can be set arbitrarily. We measure the spin-down of a microsphere released from a rotating electric field, the harmonic motion of the dipole relative to the instantaneous direction of the field, and the phase lag between the driving electric field and the dipole moment of the MS due to drag from residual gas. We also observe the gyroscopic precession of the MS when the axis of rotation of the driving field and the angular momentum of the microsphere are orthogonal. These observations are in quantitative agreement with the equation of motion. The control offered by the electrical drive enables precise measurements of microsphere properties and torque as well as a method for addressing the direction of angular momentum for an optically levitated particle.  

\end{abstract}

\maketitle

\section{INTRODUCTION}

The ability to manipulate microscopic objects has found important applications in science and technology. The interest in optical levitation of dielectric objects in vacuum, pioneered by Ashkin and Dziedzic~\cite{Ashkin:1970, Ashkin:1971}, has grown in recent times, with applications in precision measurements~\cite{Geraci:2010, Moore:2014, Rider:2016, Ranjit:2016, Arvanitaki:2013, Goldwater:2018}, surface science~\cite{Blakemore:2018} and quantum technology~\cite{Li:2011, Chang:2010, Romero-Isart:2011, Romero-Isart_2011_2, Millen:2015, Fonseca:2016, Slezak:2018, Jain:2016, Rahman2017}. The suggestion was made in~\cite{Ashkin:1970} that the rotational degrees of freedom (DOFs) of an optically levitated dielectric microsphere (MS) could be manipulated using the angular momentum in circularly polarized light. This has been realized in more recent times~\cite{Ahn:2018, Monteiro:2018, Reimann:2018, Kuhn:17,Arita:2011, Arita:18, Arita:2013, He:1995, Bishop:2004, Bennett:2013, Hoang:2016, He:1995, Mazilu:2016, Kuhn:17, Li:18, Rashid:2018}.

Here we present a technique for manipulating the rotational DOFs of an optically levitated MS by using electric fields to apply a torque to the $\lvert \vec{d} \rvert \sim 100~e\,\mu$m permanent electric dipole moment~\cite{Rider:2016} found in 2.4-$\mu$m radius silica MSs grown using the St\" ober process~\cite{Stober:1968}. The orientation of the dipole moment follows the orientation of the driving field so that the angular velocity, $\boldsymbol{\omega}_{\text{MS}}$, can be set in both magnitude and direction~\cite{Nagornykh:2017}. Using the control afforded by electric torques, we observe the spin down of a MS suddenly released from a rotating electric field, measure the MS dipole moment by tuning the libration frequency, test the relationship between residual gas pressure and drag, and induce gyroscopic precession by rapidly changing the electric field axis of rotation. Because the electric field and dipole moment are both known in this system, precise quantitative measurements are possible.     

 \section{EXPERIMENTAL TECHNIQUES AND PRINCIPLES}

The rotational response of a trapped MS, including an applied electric field, is described by the equation of motion: 
\begin{equation}
\label{eq:eoml}
\dot{\vec{L}} = \mathbfcal{T} = \vec{d}\times\vec{E} - \frac{\beta}{I}\vec{L} + \mathbfcal{T}_{\text{opt}},     
\end{equation}

\noindent where $\mathbfcal{T}$ is the total torque, $\vec{d}$ is the electric dipole moment, $\vec{E}$ is the electric field, $\vec{L}$ is the angular momentum related to the angular velocity by $\boldsymbol{\omega}_{\text{MS}}= \vec{L} /I$, $\beta$ is the rotational damping coefficient, $I$ is the moment of inertia, and $\mathbfcal{T}_{\text{opt}}$ is the optical torque. The part of the optical torque which does not average to zero over a rotation is generally negligible compared to the electric torques used here \cite{Friese:1998}. 

The angular velocity and the rotational phase of the MS are measured optically. As the MS rotates, it couples some of the incident linearly polarized optical power, $\mathcal{P}_{0}$, into the cross-polarized optical power, $\mathcal{P}_{\bot}$, according to 
\begin{equation}
\label{eq:cross_polarization}
    \mathcal{P}_{\bot} = \mathcal{P}_{0}\sin{(\eta/2)}^2\sin{\phi}^{2},
\end{equation} 
where $\eta$ is the phase retardation between the fast and slow axes, and $\phi$ is the angular displacement of the MS relative to an origin in which the fast axis is aligned with the incident polarization~\cite{Jones:41}. The $\sin{\phi}^{2}$ term implies that the phase of the MS is encoded as a modulation of $\mathcal{P}_{\bot}$, at twice the rotation frequency, $\omega_{\text{MS}}$.

The optical trap is identical to that described in Ref.~\cite{Rider:2018} with the addition of polarization optics to measure the cross-polarized light, $\mathcal{P}_{\bot}$. One polarizing beam splitter (PBS) is inserted before the trap to define the linear polarization of the incident light, and a second PBS is placed after the trap to extract $\mathcal{P}_{\bot}$ and measure the rotational phase of the MS. The remainder of the optical system, described in Ref.~\cite{Rider:2018}, is capable of stabilizing the optical trap at high vacuum.  

The residual gas pressure is controlled and measured between ${\sim} 2\times 10^{-6}$ and $1$~mbar. The vacuum pressure is tuned by introducing or removing N$_{2}$ gas and is measured by a cold-cathode gauge for pressures below 10$^{-4}$~mbar, a capacitance manometer for pressures between 10$^{-4}$ and 10$^{-2}$~mbar, and a Pirani gauge for pressures between 10$^{-3}$ and 1~mbar. The cold cathode gauge is found to affect the charge of the MS, so it is only used to measure the ${\sim} 2\times10^{-6}$~mbar base pressure of the vacuum system after an experiment. The capacitance manometer does not cover the full range of vacuum pressures, so the Pirani gauge is calibrated against the more accurate capacitance manometer, where there is overlap. This system is capable of measuring the pressure to an accuracy of 10\% for N$_{2}$ over the range of interest~\cite{kurt_lesker, MKS}.

\begin{figure}[t!]
\includegraphics[width=1.\columnwidth]{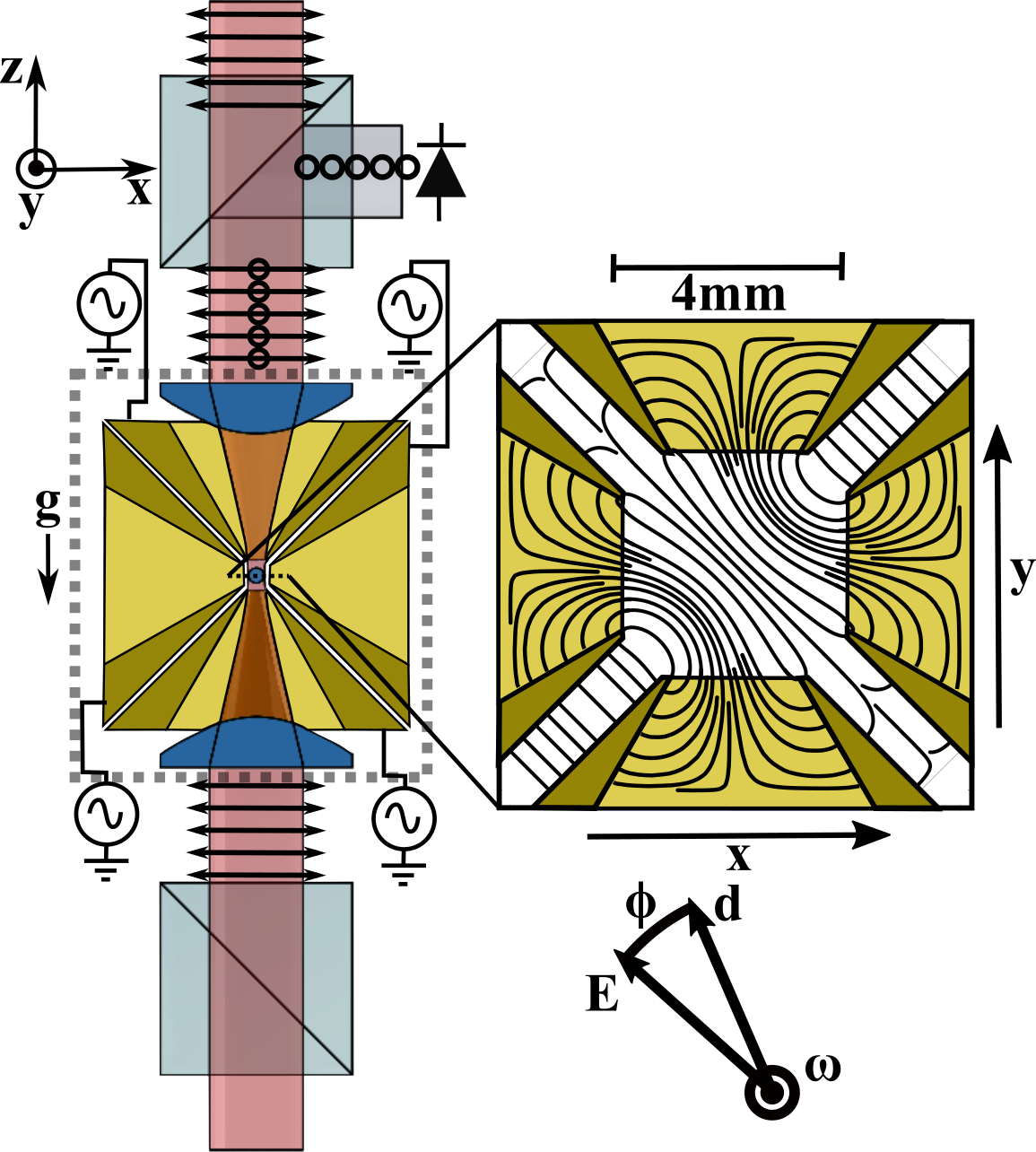}
\caption{Cross section of the electrodes and polarization optics. Each electrode is a truncated pyramid with a cone bored in the back to allow optical and mechanical access to the trapping region. The MS is levitated by an optical system identical to that described in~\cite{Rider:2018}, with the addition of the polarization optics used to measure the rotational state of the MS. The dashed lines denote the components inside the vacuum chamber. The voltages on each of the six electrodes around the trapping region are driven to exert arbitrary torques on the MS's permanent dipole moment. The detail to the right shows a slice of the electric field streamlines produced by the electrode geometry with the electrodes on the top and the left at $+V$, while those on the bottom and on the right are at $-V$. The convention used to define $\phi$ is also illustrated to the bottom right.} 
\label{fig:spinning_apparatus}
\end{figure}

The trapping region of the apparatus is illustrated in Fig.~\ref{fig:spinning_apparatus}, which shows a cross section of the truncated pyramidal electrodes used to apply torquing electric fields to the MS. A cone is bored into the back of each pyramidal electrode to allow optical and mechanical access to the trapping region. The six electrodes define a cubic trapping cavity 4~mm on a side. Each electrode is connected to a high-bandwidth, high-voltage amplifier driven by a digitally synthesized analog signal. This apparatus is capable of producing arbitrary three-dimensional electric fields up to 100~kV/m in magnitude at frequencies as high as 1~Mrad/s, which limits the rotation frequencies achieved here. To produce a spinning electric field, a sinusoidal voltage is applied to a set of four electrodes in a plane, with a phase offset of $\pi/2$ between successive electrodes. A finite element analysis is used to calculate the electric field. It is found that the $x$ component of the field in the center of the trap, $E_{x}$, is well approximated by $E_x = 0.66 (\Delta V_x / \Delta x)$, where $\Delta V_x$ is the potential difference across a pair of electrodes, and $\Delta x = 4$~mm is the electrode separation. The same statement applies to $E_{y}$ and $E_{z}$.

\begin{figure}[b!]
\includegraphics[width=1.\columnwidth]{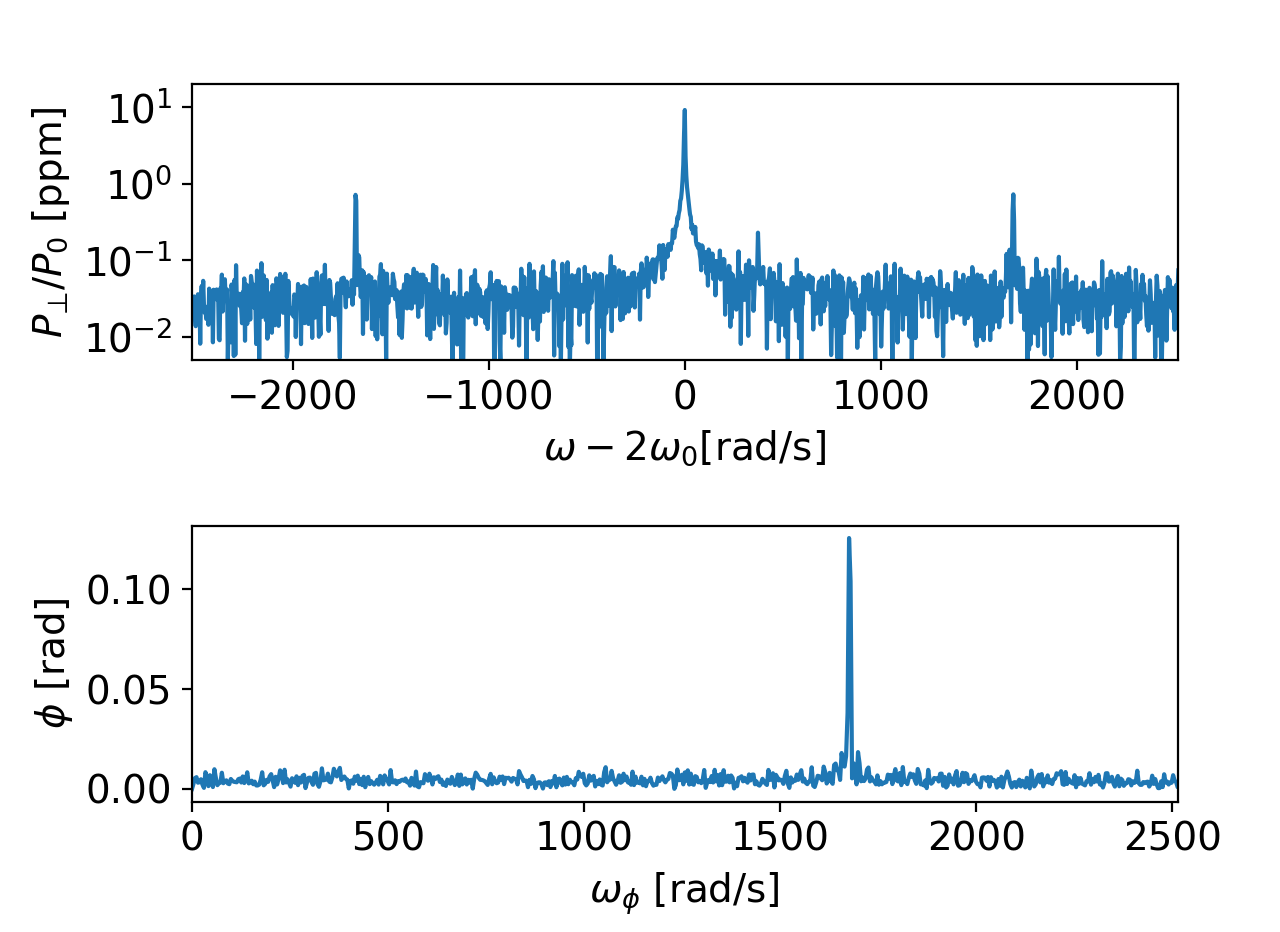}
\caption{Top: typical amplitude spectrum of $P_{\bot}$ for a MS prepared in a state of angular momentum pointing along the $\hat{\vec{z}}$ direction. The MS is spinning with $\omega_{MS} = 100\pi$~krad/s, driven by an electric field with $E = 27$~kV/m. The prominent line with sidebands are signatures of the MS rotation, with the position of the central line at twice the drive frequency. The sidebands are phase modulation of the rotation frequency, as shown in the bottom panel.}
\label{fig:rotation_spectra}
\end{figure}

Before performing a measurement, the MS is discharged as described in Refs.~\cite{Moore:2014,Blakemore:2018, Frimmer:2017}. In addition, the MS is prepared in a state of known angular momentum and rotational phase by dissipating any initial angular velocity using 0.1~mbar of N$_{2}$ gas. An electric field rotating at $\omega_0 = 2 \pi$~rad/s with $E = 41$~kV/m is then turned on to align $\vec{d}$ with $\vec{E}$. The chamber pressure is reduced to the base pressure of the vacuum system and $\omega_0$ is increased at a rate of $300$~rad/s$^{2}$ to the desired rotation frequency. Changes in the rotational dynamics in response to changes in the electric field magnitude and direction, as well as the damping coefficient, $\beta$, can then be observed. The data presented here were collected with one specific MS. Similar qualitative behavior is observed with other MSs, and the measured dipole moment is representative of the population.   

\section{OBSERVATIONS AND DYNAMICS}

A typical amplitude spectrum of $\mathcal{P}_{\bot}$, in the region around twice the drive frequency, with $\omega_{0} = $100$\pi$~krad/s is shown in the top panel of Fig.~\ref{fig:rotation_spectra}. A clear peak which follows the frequency, $\omega_{0}$, of the electric driving signal is observed at $2\omega_0$. The $\simsym \, 10$~ppm amplitude modulation caused by rotation of the MS implies that the relative phase retardation of this MS is $\eta \sim 10^{-2}$. The prominent sidebands are caused by harmonic oscillation of the dipole about the electric field, which can be seen by demodulating the phase of $\mathcal{P}_{\bot}$ relative to the electric field carrier signal as shown in the bottom panel of Fig.~\ref{fig:rotation_spectra}.

\subsection{Release from a spinning field}

\begin{figure}[t!]
\includegraphics[width=1.\columnwidth]{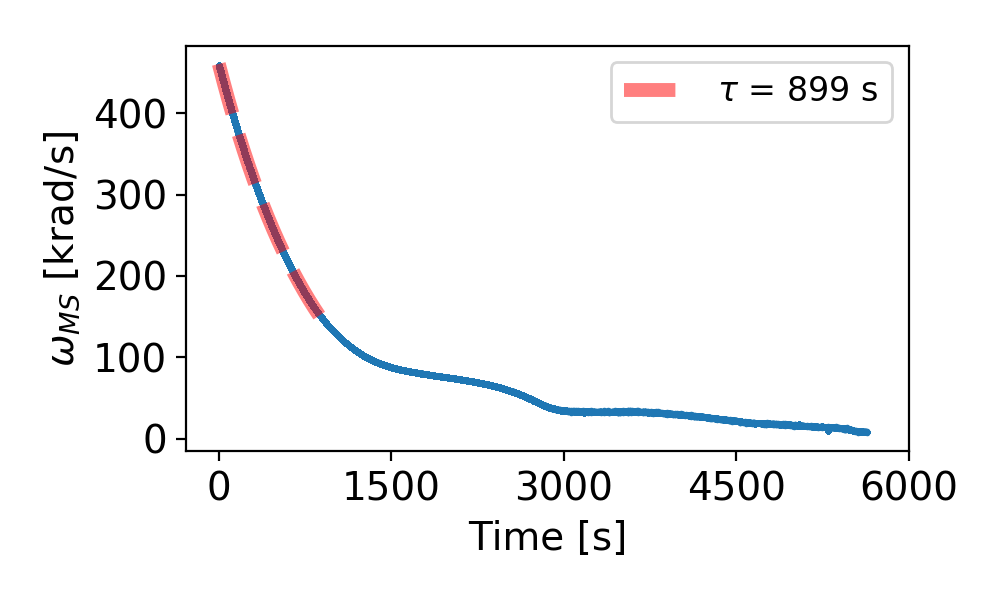}
\caption{Time evolution of $\omega_{\text{MS}}$ after the driving electric field is switched off. For $\omega_{\text{MS}}\gtrsim 150$~krad/s the angular velocity exhibits an exponential decay. For $\omega_{\text{MS}} \lesssim 150$~krad/s the dynamics are modified by torque that could be explained by a $\simsym \, 100$~ppm degree of ellipticity in the 1~mW trapping beam and the $\eta \sim \, 10^{-2}$ phase retardation of the MS.}
\label{fig:spindown}
\end{figure}

After initializing the MS with a definite angular momentum and phase, the conditions can be changed to observe different solutions of the equation of motion. The simplest solution occurs when the drive electric field initially rotating about the trapping beam axis is switched off, so that only the drag term, $-(\beta/I)\vec{L}$, remains in Eq.~\ref{eq:eoml} and the initial angular momentum decays according to 
\begin{equation}
\label{eq:decay}
  \vec{L}(t) = e^{-t/\tau}\vec{L}(0). 
\end{equation}
Here, the damping time $\tau$ is related to the damping coefficient by $\tau = I/\beta$. This decay is illustrated in Fig.~\ref{fig:spindown} at the base pressure of the vacuum system ($2\times10^{-6}$~mbar). For the first 1000~s, $\omega_{\text{MS}} \gtrsim 150$~krad/s, the drag torque dominates, and the data are well modeled by an exponential decay. The behavior beyond 1000~s can be attributed to an optical torque, $\mathbfcal{T}_{\text{opt}} \sim 10^{-23}$~Nm. 

The average optical torque on a birefringent particle is given approximately by \begin{equation}
    \mathbfcal{T}_{\text{opt}} \approx \frac{\mathcal{P}}{\omega_{\text{opt}}}\big[1 - \cos{(k r \Delta n)}\sin{2 \phi}\big],
\end{equation} 
where $\mathcal{P}$ is the $\simsym 1$~mW trapping beam power, $\omega_{\text{opt}} \simsym 10^{15}$~rad/s is the optical frequency, $k \simsym 2\pi / (1$~$\mu$m) is the wavenumber, $r = 2.35$~$\mu$m is the radius of the mircosphere, $\Delta n$ is the birefringence, and $\phi$ is the degree of ellipticity~\cite{Friese:1998, Laporta:2004}. The terminal angular velocity of the microsphere can then be explained by a fluctuating $\simsym 100$~ppm degree of ellipticity in the trapping beam.

\subsection{Libration}
\begin{figure}[b!]
\includegraphics[width=1.\columnwidth]{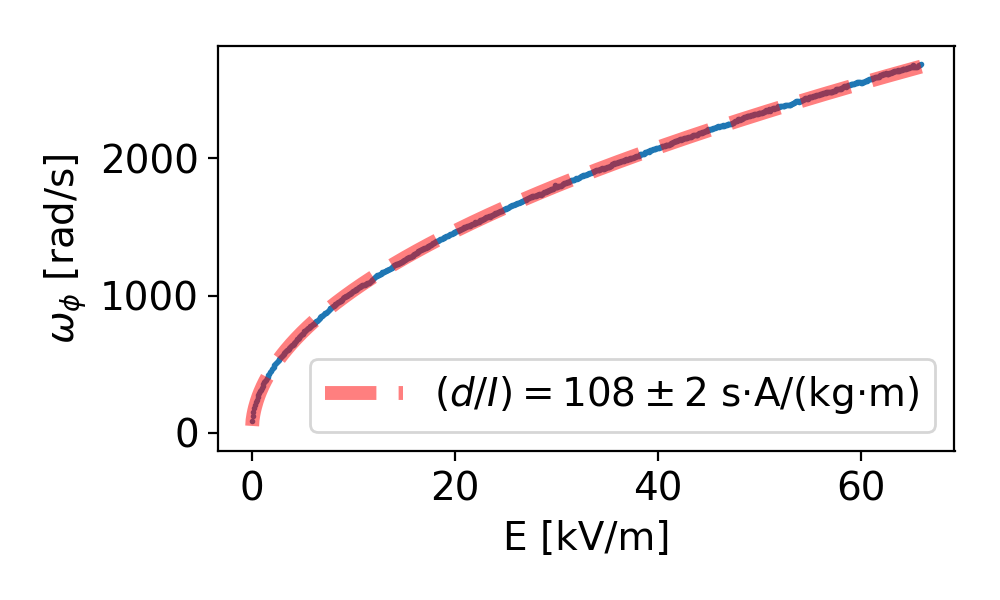}
\caption{Harmonic oscillation frequency, $\omega_{\phi}$, versus driving electric field amplitude, $E$, for a MS spinning at $\omega_{\text{MS}} = 100\pi$~krad/s at a pressure of $2\times 10^{-6}$~mbar. The data is fit to Eq.~\ref{eq:omega_phi}, obtaining $(d/I) = 108 \pm 2$~s\,A/(kg\,m) and $d = 127 \pm 14~e\,\mu$m.}
\label{fig:sideband_fit}
\end{figure}

The simplest dynamics with the electric dipole interacting with a rotating electric field can be analyzed in the case where the electric field is rotating about a fixed axis with the dipole lying in the same plane as the electric field. In the frame corotating with $\vec{E}$, Eq.~\ref{eq:eoml} reduces to an equation of motion for $\phi$, the angle between the electric field and the dipole,
\begin{equation}
    \ddot{\phi} = -\omega_{0}\Omega\sin{(\phi)}-\frac{1}{\tau}(\omega_{0}+\dot{\phi}),
\end{equation}
\noindent where
\begin{equation}
\Omega \equiv d \, E/(I \omega_{0}). 
\end{equation}
For sufficiently low damping, $\tau \Omega >1$, this equation has an equilibrium solution,
\begin{equation}
\label{eq:phi_eq}
    \phi_{\text{eq}} = -\arcsin{\left( \frac{1}{\tau \Omega} \right)} = -\arcsin{\left( \frac{\beta \omega_{0}}{d \, E} \right)},
\end{equation}
\noindent and can be linearized to give harmonic oscillation at the frequency
\begin{equation}
\label{eq:omega_phi}
    \omega_{\phi} = \sqrt{\cos{(\phi_{\text{eq}})}\omega_{0}\Omega} = \sqrt{\cos{(\phi_{\text{eq}})}\frac{E d}{I}}.
\end{equation}
This results in the sidebands shown in Fig.~\ref{fig:rotation_spectra}. The dependence of $\omega_{\phi}$ on the magnitude of the driving electric field, $E$, is well modeled by Eq.~\ref{eq:omega_phi}, as shown in Fig.~\ref{fig:sideband_fit}. The equilibrium phase lag $\phi_{\text{eq}}$ may be neglected because $\tau \Omega \gg 1$ at the base presssure of the vacuum system. The fit shown in Fig.~\ref{fig:sideband_fit} extracts the ratio $d/I$, which can be used to determine the dipole moment, $d$, if the MS is assumed to be a uniform sphere with the radius, $r_{MS} = 2.35 \pm 0.02$~$\mu$m and mass $M_{MS} = 85\pm9$~pg, measured for this lot of MS in Ref.~\cite{Blakemore:2018_2}. This procedure gives $I = (1.9 \pm 0.2) \times 10^{-25}$~kg\,m$^2$, which implies $d = 127\pm14~e \, \mu$m, in agreement with Ref.~\cite{Rider:2016}. The ability to measure the dipole moment from the frequency of harmonic oscillation enables precise measurements of torques on an optically levitated particle in high vacuum by balancing an unknown torque against an electric torque.

\subsection{Drag from residual gas}

Once the dipole moment is known, it is possible to measure the drag due to residual gas and verify models of dissipation for optically levitated particles. The equilibrium phase lag $\phi_{\text{eq}}$ between $\vec{d}$ and $\vec E$ is expected to increase with the pressure $P$, as the drag from the gas increases. In the molecular flow regime, the damping coefficient, $\beta$, can be written as $\beta = k\,P$, where $k$ is a constant that depends on the geometry of the MS, as well as the temperature and species of residual gas~\cite{Cavalleri2010, Martinetz:2018}. The argument to the $\arcsin$ in Eq.~\ref{eq:phi_eq} can then be parameterized by 
\begin{equation}
\label{eq:PoverE}
    \frac{\omega_{0}\beta}{d \, E} = \frac{\omega_{0} k}{d \, E} P = \frac{P}{P_{\pi/2}},
\end{equation}
\noindent where $P_{\pi/2} \equiv d \, E/(\omega_{0}k)$ is the pressure at which $\phi_{\text{eq}} \rightarrow -\pi/2$, where the MS rotation loses lock with the driving electric field (in practice fluctuations cause the MS to lose lock before $\phi_{\text{eq}}$ reaches $-\pi/2$). 

\begin{figure}[t!]
\includegraphics[width=1.0\columnwidth]{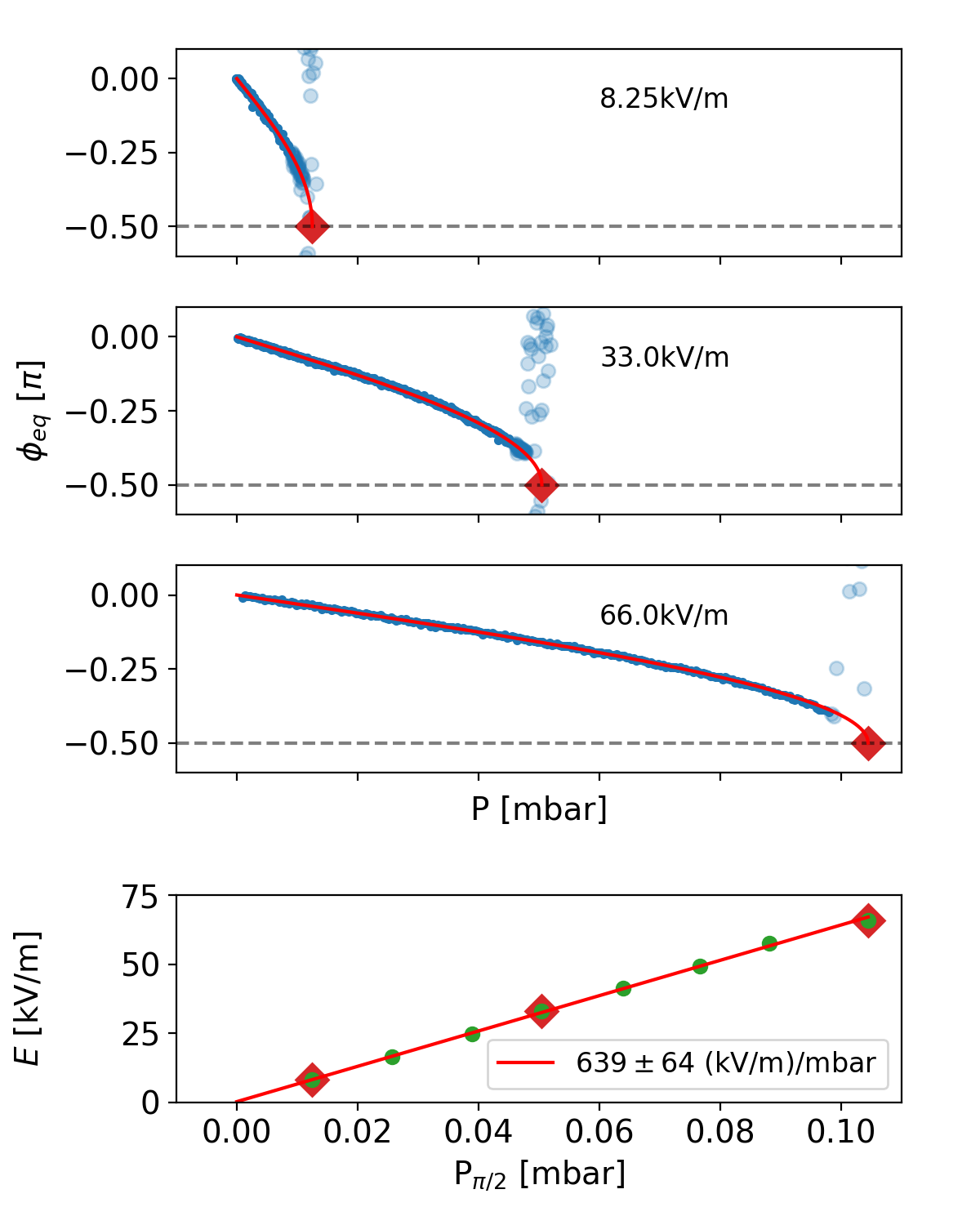}
\caption{Top three panels: Equilibrium phase, $\phi_{eq}$, versus chamber pressure for several magnitudes, $E$, of the driving field with $\omega_{MS} = 100\pi$~krad/s. For each value of $E$, $\phi_{eq}$ increases until the MS loses lock with the field, and the phase becomes random. For each $E$, a fit to Eq.~\ref{eq:phi_eq} (with the argument in Eq.~\ref{eq:PoverE}) is shown in red. $P_{\pi/2}$ is identified by a red diamond. Bottom panel: The linear relationship between $P_{\pi/2}$ and $E$, with additional $E$ included. The slope of the fit of $E$ vs $p_{max}$ is $639\pm64 $~(kV/m)/mbar. Assuming the dipole moment measured from the frequency of small oscillations, this gives $k = \beta/P = (4.1\pm 0.6) \times 10^{-25}$~m$^{3}$s, which is consistent with $k = \beta/P = 3.4 \times 10^{-25}$ ~m$^{3}$s predicted in Refs.~\cite{Cavalleri2010, Martinetz:2018}.}

\label{fig:phase_vs_p}
\end{figure}

This behavior is shown in the three top panels of Fig.~\ref{fig:phase_vs_p} for three different amplitudes of the driving field. It is evident that the unlocking pressure depends on the field amplitude and that after losing lock $\phi_{\text{eq}}$ becomes random. $P_{\pi/2}$ can be extracted from a fit for each field amplitude, as plotted (with additional values of the field) in the bottom panel of Fig.~\ref{fig:phase_vs_p}. The linear relationship $k = d\,E/(\omega_{0}P_{\pi/2})$ indicates that the dissipation is proportional to the residual gas pressure and there are no significant additional sources of dissipation at pressures $\gtrapprox 10^{-2}$~mbar. 

 The fit reports $k= (4.1\pm0.6)\times10^{-25}$~m$^3$s, assuming $d = 127\pm14~e{\cdotp}\mu$\,m.  This is consistent with the value $k = 3.4\times 10^{-25}$~m$^3$s predicted in Refs.~\cite{Cavalleri2010, Martinetz:2018} for a $2.35$-$\mu$m-radius MS in thermal equilibrium with 300~K N$_{2}$ gas. No evidence for increased dissipation due to an elevated MS temperature or surface roughness is observed. This measurement of $k$ can be used to infer the base vacuum pressure in the vicinity of the MS for the spin-down time measured in Fig.~\ref{fig:spindown} from $P = I/(k \tau)$. This give $P = (5.1 \pm 0.9)\times 10^{-6}$~mbar, which is roughly a factor of two greater than the pressure measured after the experiment with the cold cathode gauge. This discrepancy could be due to another source of dissipation which becomes significant at low pressures, inaccuracy of the cold cathode gauge, or a real pressure difference between the residual gas pressure in the trap and the cold cathode gauge at the end of the measurement.

\subsection{Gyroscopic precession}

When the electric field rotation axis and the angular momentum are not aligned, the dynamics are complex and depend sensitively on the initial conditions. For $\Omega \ll \omega_{0}$, and a field $\vec{E}(t) = E\,[\cos{(\omega_{0}t)}\hat{\vec{z}} -\sin{(\omega_{0}t})\hat{\vec{y}}]$, rotating about the $\hat{\vec{x}}$ axis, an approximate solution is given by
\begin{equation}
\label{eq:procession_L}
    \vec{L}(t) = L\,\left\{ \cos{\left[ (\Omega/2) t\right]} \, \vec{\hat{z}} + \sin{ \left[ (\Omega/2) t\right]} \, \hat{\vec{y}} \right\}
\end{equation}
and

\begin{equation}
\label{eq:procession_p}
\begin{split}
    \vec{d}(t) = - d \, \Big\{ &\cos{ \left[ (\omega_0 + \Omega/2) t \right]} ~ \hat{\vec{x}} \\
    & + \sin{\left[ \left( \omega_{0} + \Omega/2 \right) t\right]}\cos{\left[ (\Omega/2) t \right]} ~  \hat{\vec{y}} \\
    & - \sin{\left[ \left( \omega_{0} + \Omega/2 \right) t\right]}\sin{\left[ (\Omega/2) t \right]} ~ \hat{\vec{z}} \Big\},
\end{split}
\end{equation}

\noindent in the absence of dissipation.
\begin{figure}[t!]
\includegraphics[width=1.0\columnwidth]{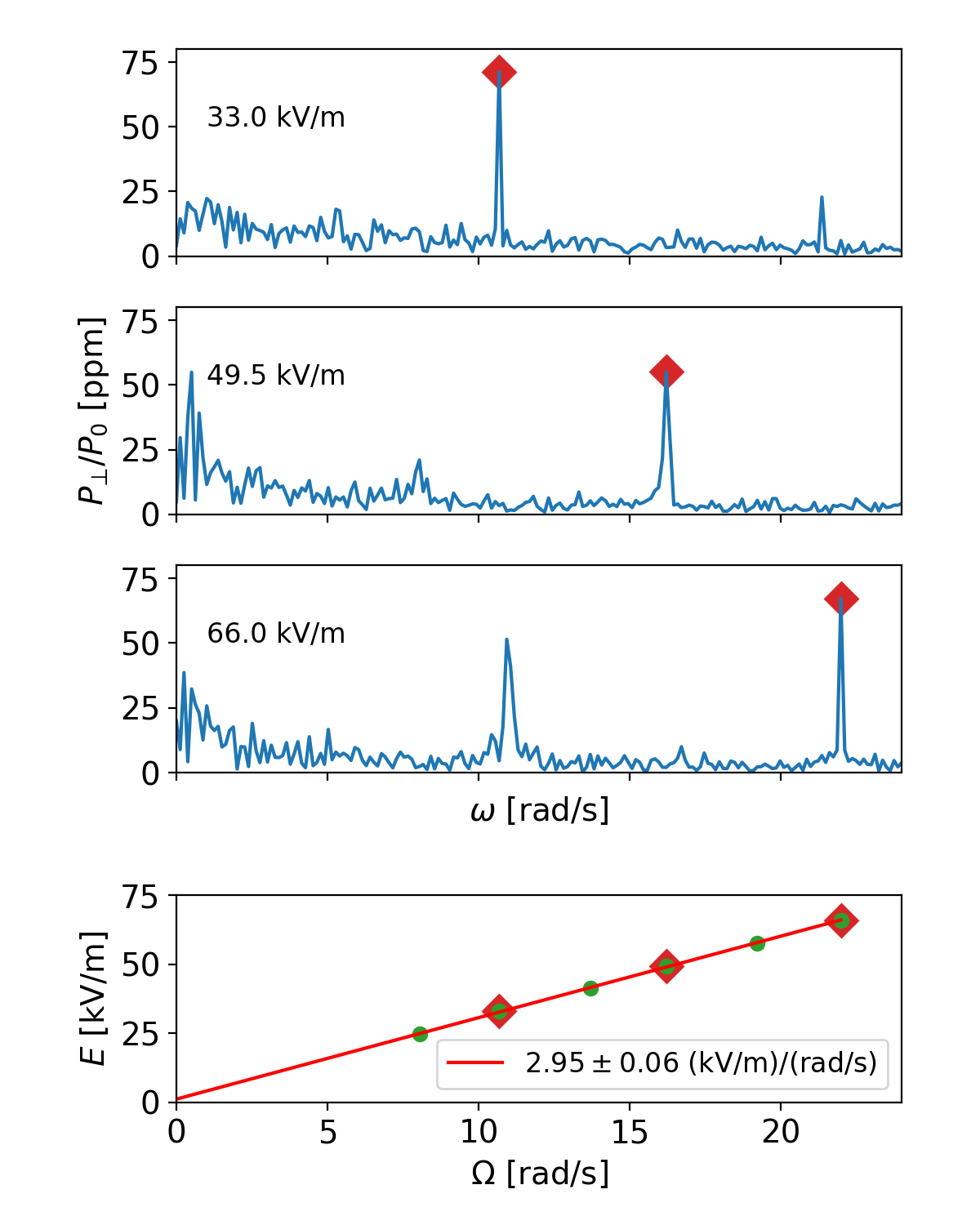}
\caption{Top three panels: Spectra of the cross-polarized light intensity, $\mathcal{P}_{\bot}$, for a MS precessing about the $\hat{\vec{x}}$ axis while spinning at $\omega_{\text{MS}} = 100 \pi$~krad/s). The modulation of the cross-polarized light occurs predominantly at twice the precession frequency, denoted by red diamonds. Bottom panel: $\Omega$ for different $E$. The slope of the fit (red line) provides $(d/I) = \omega_{0}\Omega/E = 106 \pm 2$~(s\,A)/(kg\,m), which is consistent with the measurement of $d/I$ from the frequency of small oscillations.}
\label{fig:precession}
\end{figure}

The angular momentum vector $\vec{L}$ undergoes left-handed precession about the electric field axis of rotation at an angular frequency $\Omega/2$. The factor of $1/2$ is due to the fact that the torque is averaged over a rotation of the MS. This behavior is observed as a low-frequency modulation of the cross-polarized light most prominently at twice the precession frequency. The amplitude spectra for a precessing MS driven by a range of electric fields magnitudes are shown in Fig.~\ref{fig:precession}, in which the MS is spinning at $\omega_{0} = 100 \pi~$krad/s. The low-frequency amplitude modulations of the cross-polarized light are larger than those for rotation about the trapping beam axis because of the coupling with the feedback used to stabilize the microsphere at low frequencies. As expected, the precession frequency is proportional to the magnitude of the driving electric field and the slope of $\Omega$ vs $E$, which implies $(d/I) = 106 \pm 2$~(s\,A)/(kg\,m), is consistent with the value of $(d/I) = 108 \pm 2$~(s\,A)/(kg\,m) from the measurement of libration. 

\section{CONCLUSION}

We have demonstrated electrically driven rotation of optically levitated particles. Although it would be technically challenging to reach the $\simsym1$~GHz rotation frequencies achieved with optical rotation~\cite{Reimann:2018}, electrically driven rotation offers precise control over the direction and magnitude of the torque, making the quantitative measurements presented here possible. This is in contrast to optically driven rotation, where the coupling to the optical torque depends on the geometry and birefringence of the trapped particle, as well as the exact optical properties of the laser beam used to apply the torque. The ability to measure and control system parameters has implications for the field of optical levitation and manipulation.      

The libration of an optically levitated dipole trapped in an electric field is a degree of freedom that can be exploited for cooling and precision measurement. Libration of a dipole trapped in a rotating electric field can be cooled by simple phase modulation of the electric field driving the rotation. MSs made from materials with larger dipole moments could also be used to couple to and measure oscillating electric fields near the rotation frequency. After the dipole coupling to an electric field is measured from the libration frequency, torques measured from the phase lag of the rotation can be calibrated into physical units.  

Background forces arising from electric fields coupling to a MS dipole moment have been the limiting factor in several measurements with optically levitated MSs~\cite{Moore:2014, Rider:2016}. The techniques presented here provide new tools to control and measure the orientation of a MS's dipole moment, which could be used to mitigate these background forces by averaging over a rotation of the dipole. This would not be possible with optical rotation because the dipole is not guaranteed to be orthogonal to the axis of rotation. 

The measurement of drag torque versus residual gas pressure demonstrates a technique for precise quantitative torque measurements with optically levitated MSs and validates a model for the drag on a levitated particle in the molecular flow regime. These techniques could be used to measure the optical properties of a levitated particle by balancing electric and optical torques.

Finally, the demonstration of gyroscopic precession induced by misalignment between the angular momentum and the electric field rotation axes demonstrates a technique for measuring the orientation of the angular momentum for an electrically driven MS. This opens the way to applications in the area of microscopic gyroscopes, with rotational drag which may be engineered to be at or beyond the level of current mechanical gyroscopes and the possibility of sensing torques at microscopic length scales.

\begin{acknowledgments}

We would like to thank J.~Fox for discussions on the readout electronics, R.G.~DeVoe and A.~Fieguth for carefully reading this manuscript, and the Moore group for general discussions related to trapping microspheres. This work was supported, in part, by NSF Grants No. PHY1502156 and No. PHY1802952, ONR Grant No. N00014-18-1-2409 and the Heising-Simons Foundation. AK acknowledges the partial support of a William~M. and Jane~D. Fairbank Postdoctoral Fellowship of Stanford University. NP acknowledges the partial support of the Koret Foundation. Microsphere characterization was performed at the Stanford Nano Shared Facilities (SNSF), which is supported by the National Science Foundation as part of the National Nanotechnology Coordinated Infrastructure under award ECCS-1542152.
\end{acknowledgments}

\bibliographystyle{apsrev4-1}
\bibliography{spinning_MS}

\end{document}